\documentclass[aps,prb,preprint,showpacs,superscriptaddress,floatfix]{revtex4-2}

\usepackage{graphicx}
\usepackage{float}
\usepackage{amsmath}
\usepackage{xcolor}
\usepackage{bm} 
\usepackage{hyperref}

\renewcommand{\vec}[1]{\bm{#1}} 

\usepackage{array}
\newcolumntype{P}[1]{>{\centering\arraybackslash}p{#1}}

\begin{document}

\title{Lithium intercalation in MoS$_2$ bilayers and implications for moir\'e flat bands}
\author{Zheyu Lu} 
\altaffiliation[Current address: ]
{Department of Physics, University of Calilfornia, Berkeley, California, 94720, USA.}
\affiliation{Department of Physics, University of Science
and Technology of China, Hefei, Anhui, 230026, China.}
\author{Stephen Carr}
\affiliation{Department of Physics, Harvard University, Cambridge, Massachusetts, 02138, USA.}
\author{Daniel T. Larson}
\affiliation{Department of Physics, Harvard University, Cambridge, Massachusetts, 02138, USA.}
\author{Efthimios Kaxiras}
\affiliation{Department of Physics, Harvard University, Cambridge, Massachusetts, 02138, USA.}
\affiliation{John A. Paulson School of Engineering and Applied Sciences, Harvard University, Cambridge, Massachusetts, 02138, USA.}
\date{\today}

\begin{abstract}

Intercalation of lithium atoms between layers of 2D materials can alter their atomic and electronic structure.
We investigate effects of Li intercalation in twisted bilayers of the transition metal dichalcogenide MoS$_2$ through first-principles calculations, tight-binding parameterization based on the Wannier transformation, and analysis of moir\'e band structures through an effective continuum model.
The energetic stability of different intercalation sites for Li between layers of MoS$_2$ are classified according to the local coordination type and the number of vertically aligned Mo atoms, suggesting that the Li atoms will cluster in certain regions of the moir\'e superlattice.
The proximity of a Li atom has a dramatic influence on the interlayer interaction between sulfur atoms, deepening the moir\'e potential well and leading to better isolation of the flat bands in the energy spectrum.
These results point to the usefulness for the use of chemical intercalation as a powerful means for controlling moir\'e flat-band physics in 2D semiconductors.
\end{abstract}

\maketitle

\section{Introduction}
Strongly correlated insulating behavior and unconventional superconductivity have recently been observed in magic-angle twisted bilayer graphene~\cite{cao2018correlated,cao2018unconventional}. The small twist angle between the two graphene layers creates a moir\'e pattern with a characteristic length scale much greater than the lattice constant of the individual layers. The interlayer hybridization of the two layers' Dirac-cones results in the emergence of flat bands in the low-energy band structure~\cite{rafi2011}. This reduction of the electrons' kinetic energy favors the usually weak electron-electron interactions and phonon-electron coupling in graphene, leading to twist-induced correlated behavior.

Flat bands and correlated physics have been predicted and observed in other moir\'e superlattices, such as trilayer graphene on top of hexagonal boron nitride~\cite{chen2019evidence,chen2019signatures} and twisted bilayer-bilayer graphene~\cite{Shen2019,cao2019electric,liu2019spin, Chebrolu2019, Koshino2019, Lee2019}. In addition, novel absorption peaks, interpreted as intralayer and interlayer moir\'e excitons, have been observed in different twisted bilayer transition metal dichalcogenide (TMDC) moir\'e superlattices due to enhanced electron-hole interactions~\cite{jin2019observation,seyler2019signatures,tran2019evidence,alexeev2019resonantly}. Moir\'e flat bands were predicted to form at the band edges of twisted bilayer TMDC systems~\cite{wu2018,wu2019,Naik2018,carr2018duality}, and they have recently been observed~\cite{Wang2019}. These discoveries have demonstrated how the twist angle can be a powerful tool for engineering new and interesting properties in two-dimensional van der Waals heterostructures.

Experimental control of the twist angle can be combined with other tunable perturbations common in nanomaterial experiments. For example, vertical pressure~\cite{yankowitz2019tuning,Carr2018pressure,Chittari2018} and external strain~\cite{Bi2019} have been investigated as additional tools to realize flat bands in a wider range of geometries and twist angles. At the same time, intercalation of Li atoms has been used to electrochemically dope the layers in van der Waals heterostructures~\cite{bediako_heterointerface_2018,larson2018lithium}, and has been predicted to enhance the interlayer coupling in the AA-stacked regions of twisted bilayer graphene~\cite{larson2020effects}. In the present work we explore the effects of lithium intercalation in various untwisted, local stacking arrangements of two layers of MoS$_2$, demonstrating significant amplification of the interlayer interactions caused by nearby lithium intercalants. From the calculated changes in local electronic structure, we present a continuum model for the band structure of twisted bilayer MoS$_2$ at small twist angles. The intercalants enhance the moir\'e potential, leading to better isolation of the flat bands in the energy spectrum.

The paper is organized as follows: In Sect.~\ref{methods} we describe the density functional theory (DFT) and Wannier transformation formalism used to extract \emph{ab initio} parameters used in our modeling. The crystal structure of twisted bilayer MoS$_2$ is described in Sect.~\ref{energetics}, along with our results on the energetics of Li intercalation between the layers. In Sect.~\ref{interlayer} we study the effect of Li intercalants on the interlayer coupling using both tight-binding and continuum models. We present the moir\'e band structure, including the effects of Li atoms, in Sect.~\ref{flat-bands}. Our conclusions are presented in Sect.~\ref{conclusion}.

\section{Computational Methods}\label{methods}
DFT calculations were performed using the Vienna \textit{Ab initio} Simulation Package (\texttt{VASP}) \cite{kresse1996efficient,kresse1996efficiency}. The interaction between ionic cores and valence electrons is described by pseudopotentials of the projector augmented wave type. We used the SCAN meta-GGA exchange correlation functional~\cite{sun2015strongly}, along with the rVV10 van der Waals functional~\cite{peng2016versatile}. We employed a slab geometry to model double layers with a 22 \AA{} vacuum region between periodic images to minimize the interaction between slabs. The crystal structure was relaxed until Hellmann-Feynman forces were smaller in magnitude than 0.01 eV/\AA{} for each atom. The plane-wave energy cutoff was 350 eV with a reciprocal space grid of size 17$\times$17$\times$1 for the primitive unit cell, and grids of size 9$\times$9$\times$1, 6$\times$6$\times$1, and 4$\times$4$\times$1 for the 2$\times$2, 3$\times$3, and 4$\times$4 supercells, respectively. We calculated only unrotated geometries, and assessed the implications for moir\'e systems by sampling over different atomic registries between the layers.

To extract tight-binding parameters we transform the plane-wave DFT basis into a basis of maximally localized Wannier functions (MLWF)~\cite{Marzari2012} as implemented in the Wannier90 code~\cite{mostofi2008wannier90,Pizzi_2020}. For the Wannier transformation of bilayer MoS$_2$ with Li intercalants we use the seven highest valence bands and four lowest conduction bands which are composed of Mo $d$-orbitals and S $p$-orbitals~\cite{fang2015}. We do not need to include the Li $s$-orbitals because electrons from Li are primarily transferred to the surrounding layers and raise the Fermi level into the MoS$_2$ conduction bands, but do not form new $s$-bands near $E_F$. The initial projections are chosen to be the atomic $p$- and $d$-orbitals and the final converged Wannier functions remain very similar to the localized atomic orbitals. The effective Hamiltonian in the Wannier basis is interpreted as the full-range \textit{ab initio} tight-binding Hamiltonian (FTBH)~\cite{fang2015}.

In practice, MoS$_2$ is an $n$-type semiconductor \cite{Chhowalla2016} likely due to sulfur vacancies that form during fabrication \cite{Yang2019}.
Lithium intercalation adds electrons (negative charge carriers) and enhances the $n$-type doping. Thus, for prediction of transport properties we focus on the conduction band-edge, which is at the $K$ and $K'$ points of the monolayer Brillouin Zone. This band-edge has very weak spin-splitting in MoS$_2$ ($\sim$3 meV), and so we perform calculations in the absence of spin-orbit coupling. Although this choice reduces the accuracy of our electronic structure calculations, particularly around the valence $K$-point band edge, it greatly simplifies the tight-binding parametrization and the form of the twisted continuum model.

\section{Energetics of Lithium intercalation}\label{energetics}

Each layer of MoS$_2$ is formed by a triangular lattice of Mo atoms sandwiched between two triangular lattices of sulfur atoms. In the naturally occurring 2H bulk phase the S atoms of each layer surround the Mo atoms with trigonal prismatic coordination, and each consecutive layer is rotated by 180$^\circ$ from the one below. An ``aligned" bilayer can also be fabricated, where the consecutive layers have the same orientation. Here we focus on the results for a 2H bilayer; the results for the aligned case are similar. We have not studied the 1T structure, where the Mo atoms are octahedrally coordinated within each of the layers.

\begin{figure}[htbp]
    \centering
    \includegraphics[width=0.5\linewidth]{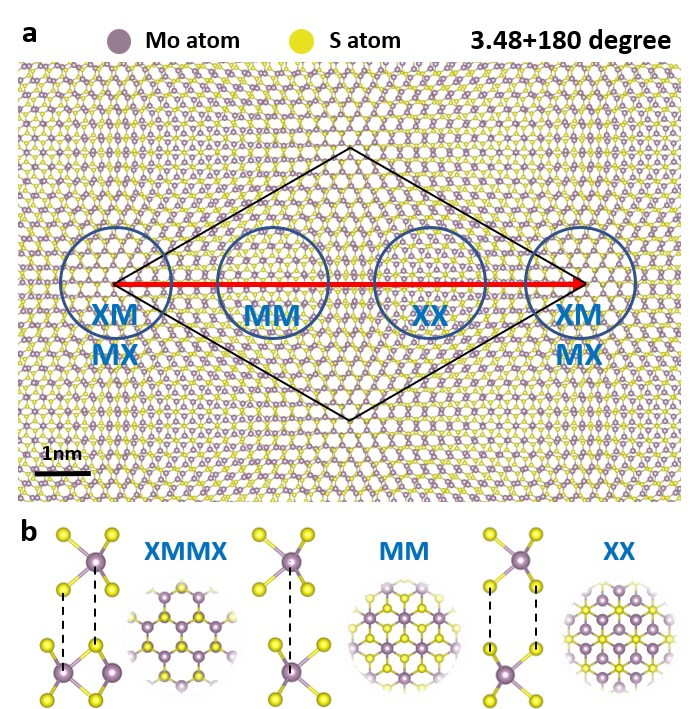}
    \caption{
    \textbf{(a)} Top view of a twisted homobilayer of MoS$_2$ in the 2H phase.
    Different local atomic configurations occur in the small twist-angle moir\'e supercell.
    The highlighted regions correspond to local atomic configurations with three-fold rotational symmetry, labeled as XMMX, MM, XX, and XMMX respectively along the diagonal direction.
    \textbf{(b)} Side and top views of the high-symmetry local configurations. XMMX stacking refers to the configuration with S of the top layer aligned with Mo of the bottom layer and Mo of the top layer aligned with S of the bottom layer; MM (XX) stacking region refers to the configuration where Mo (S) of the two layers are aligned.}
    \label{fig:1}
\end{figure}

When one layer of a MoS$_2$ bilayer has a relative twist with respect to the other layer, a large scale moir\'e pattern forms with periodicity represented by the corresponding moir\'e supercell, as shown in Fig.~\ref{fig:1}a. Within the supercell the local stacking arrangement will vary, and along the diagonal of the moir\'e supercell there are three special local stacking patterns with three-fold rotational symmetry. In graphene these three regions are referred to as AA, AB, and BA stacking, where the ``A" and ``B" labels refer to the two sublattices of the honeycomb lattice. For a 2H bilayer there are more stacking possibilities, which we label by the pairs of atoms that are vertically aligned, as shown in Fig.~\ref{fig:1}b. For a general TMDC with chemical formula MX$_2$ the 2H structure allows for XMMX, MM, and XX stacking, while the aligned structure allows for MMXX and MX stacking. In XMMX stacking a chalcogen atom of the bottom layer is directly beneath a metal atom of the top layer, and vice versa. For MM and XX the metal atoms or chalcogen atoms are vertically aligned, respectively.

\begin{figure}[htbp]
    \centering
    \includegraphics[width=0.5\linewidth]{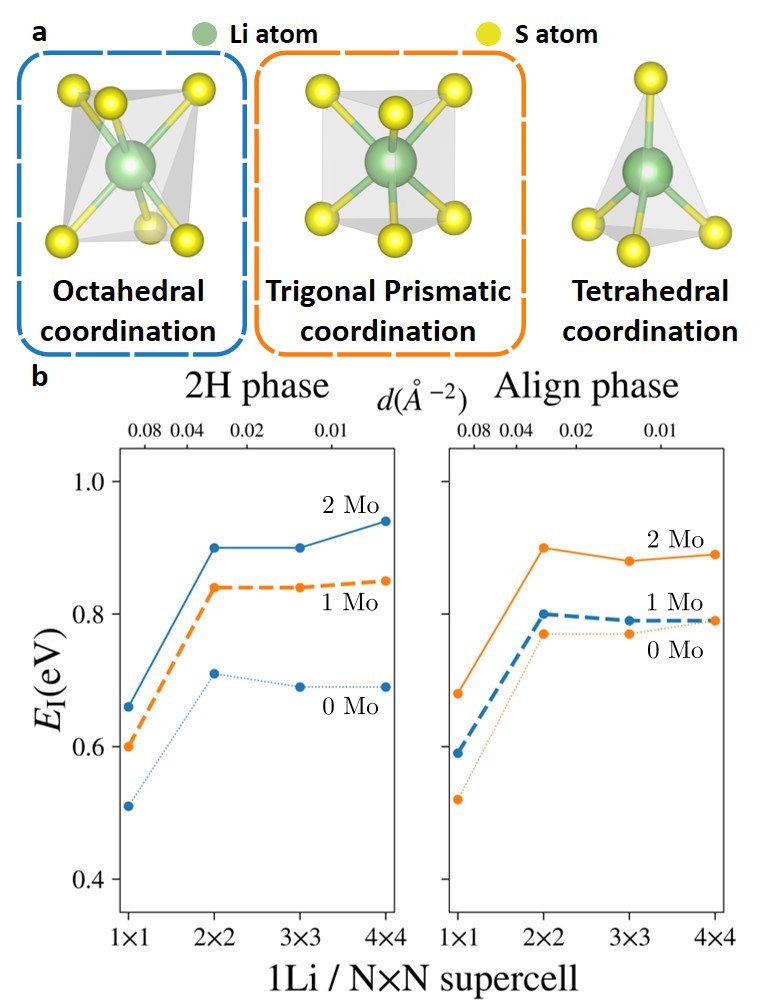}
    \caption{
    \textbf{(a)} Atomic geometries of lithium intercalation sites with different local coordination environments: octahedral, trigonal prismatic, and tetrahedral coordination with six, six, and four S atoms symmetrically arranged around the lithium atom.
    \textbf{(b)} Li intercalation energy for both the 2H and aligned phases as a function of supercell size (horizontal axis); the corresponding Li concentration, $d$, is shown on the top.
    Each curve corresponds to a specific local coordination type and number of vertically aligned Mo atoms.
    Blue and orange denote octahedral and trigonal prismatic coordination respectively.
    Solid, dashed, and dotted lines denote 2, 1, and 0 vertically aligned Mo atoms. The intercalation energy does not change significantly for concentrations below 1 lithium per 2$\times$2 supercell, $\sim$ 0.03 \AA$^{-2}$. Increasing the concentration from 0.25 Li/Mo ($2 \times 2$ supercell) to 1 Li/Mo ($1 \times 1$ supercell) has an energy cost of $\sim$200 meV.}
    \label{fig:2}
\end{figure}

Given these different local stacking arrangements in a moir\'e supercell, it is important to determine the preferred locations for Li intercalants. For a single MoS$_2$ layer, Li adsorption occurs in the hollows of the lattice formed by S atoms, either directly above a Mo atom or an empty site in the Mo lattice. When intercalating between two MoS$_2$ layers a Li atom in a sulfur hollow of one layer can experience several different arrangements of sulfur atoms from the other layer, resulting in octahedral, trigonal prismatic, or tetrahedral coordination (Fig. \ref{fig:2}a). In addition to the sulfur coordination of the Li atom, the number of vertically aligned Mo atoms is also important. Based on this argument, we label the possible intercalation sites by their sulfur coordination and number of vertically aligned Mo atoms. For the 2H bilayer there are 5 possibilities: octahedral-0, octahedral-2, trigonal-1, tetrahedral-0, and tetrahedral-1. The aligned case also has 5 intercalation sites: octahedral-1, trigonal-0, trigonal-2, tetrahedral-0, and tetrahedral-1.

The intercalation energy of lithium atoms between layers of MoS$_2$ is defined as follows~\cite{shirodkar2016}:
\begin{equation}
E_{\rm{I}} = \frac{1}{n}(E_{\rm{MoS_2}} + n E_{\rm{Li}} - E_{\mathrm{MoS}_2 + n \mathrm{Li}}).
\end{equation}
Here $E_{\rm MoS_2}$ is the energy of the empty, relaxed bilayer of MoS$_2$, $E_{\rm Li}$ is the energy of a single Li atom in bulk bcc lithium, and $E_{\mathrm{MoS}_2 + n \mathrm{Li}}$ is the energy for the bilayer containing $n$ Li atoms. $E_\mathrm{I}$ gives the decrease in the total energy of the system for each Li ion intercalated.

We calculate $E_\mathrm{I}$ for a single Li atom in a primitive cell of the bilayer, with one MoS$_2$ layer shifted (but not rotated) relative to the other layer, in order to produce the desired sulfur coordination and alignment with Mo. In order to understand the effect of Li-Li interactions, we repeated the calculations for a single Li atom in larger (still unrotated) bilayer supercells, which is equivalent to decreasing the Li concentration, $d$. For $N\times N$ supercells containing a single Li intercalant the concentration of Li is:
\begin{equation}
    d = \frac{1}{Na\times Na\times \sin{\frac{\pi}{3}}} = \frac{2\sqrt{3}}{3N^2a^2},
\end{equation}
where $a=3.19$ \AA{} is the primitive cell lattice constant.
Fig.~\ref{fig:2}(b) shows the intercalation energy for a single Li atom in each of the intercalation sites as a function of the size of the supercell. Larger supercells correspond to lower Li concentration.

For fixed Li concentration and the same number of vertically aligned Mo atoms, it is not surprising that octahedral and trigonal prismatic coordination, both with six nearest-neighbor ligands, are almost degenerate and more stable than tetrahedral coordination, which has only four nearest-neighbors. The intercalation energy, and hence stability, also increases with the number of vertically aligned Mo atoms, due to a stabilizing charge transfer between the Li and Mo atoms. From Fig.~\ref{fig:2}(b) we see that for the 2H phase the octahedral-2 location, which corresponds to MM stacking, is the most energetically favorable, followed by the trigonal-1 location (XX stacking) and then the octahedral-0 location (XMMX stacking).

For each case, $E_\mathrm{I}$ rapidly reaches a plateau with increasing supercell size (decreasing Li concentration), except for tetrahedral coordination in which case the structure is not stable, possessing negative-frequency phonon modes. The energy cost to increase Li concentration from 0.25 to 1.0 Li per bilayer unit-cell is  $\sim$200 meV, and is caused by the repulsive nature of neighboring Li$^{+}$ ions.
Because the intercalation energy does not change significantly for 2$\times$2 and larger supercells, in subsequent calculations we will use 2$\times$2 supercells where the Li ions are separated by $\sim$6.4 \AA{}. Note that for Li intercalants between twisted layers of graphene the intercalation energy is sensitive to Li-Li separations up to $\sim$15 \AA{}~\cite{larson2020effects}. But even if similar long-range interactions are present for Li atoms between MoS$_2$ layers, the primary contributions to the intercalation energy are captured already by the results of the 2$\times$2 supercells. Furthermore, as will be shown below, the effect of Li atoms on the interlayer couplings are only relevant for S-S pairs within $\sim$3 \AA{} of each Li atom, so sulfur pairs in the 2$\times$2 supercell will be influenced by at most 1 Li atom.

The Li intercalants also modify the distances between the MoS$_2$ layers.
Calculations with 1 Li atom in a $2 \times 2$ supercell indicate that Li always increases the interlayer distance, but the amount depends on its local coordination, as shown in Table~\ref{tab:interlayer_d}.
Unsurprisingly, the tetrahedral coordination shows the largest change in the interlayer distance because there is a sulfur atom directly above the intercalant.
Even in $4 \times 4$ supercells the MoS$_2$ layers remain nearly flat with height variations in the Mo atom of only $0.01$ \AA{}, but due to the lower effective Li concentration the layer separations are closer to the unintercalated values.
To obtain accurate height profiles in a moir\'e system requires a twisted supercell calculation \cite{larson2020effects}.

\begin{table}[htbp]
    \centering
    \begin{tabular}{ P{4cm} P{2cm} P{2cm} P{2cm} }
    \hline\hline
    & \multicolumn{2}{c}{$\Delta z$ (\AA{})} \\
        Geometry & 0 Li & 1 Li \\
    \hline
        2H octahedral-2    & 6.26 & 6.48 \\
        2H octahedral-0    & 6.26 & 6.55 \\
        2H trigonal-1      & 6.85 & 6.90 \\
        2H tetrahedral-1   & 6.26 & 6.86 \\
        2H tetrahedral-0   & 6.26 & 7.02 \\

    \hline
    \end{tabular}
    \caption{Average vertical distance $\Delta z$ (\AA{}) between Mo atoms in the two layers in a $2 \times 2$ bilayer supercell with and without a Li intercalant.}
    \label{tab:interlayer_d}
\end{table}

\section{Effect of lithium on interlayer couplings}\label{interlayer}

\begin{figure}[htbp]
    \centering
    \includegraphics[width=0.5\linewidth]{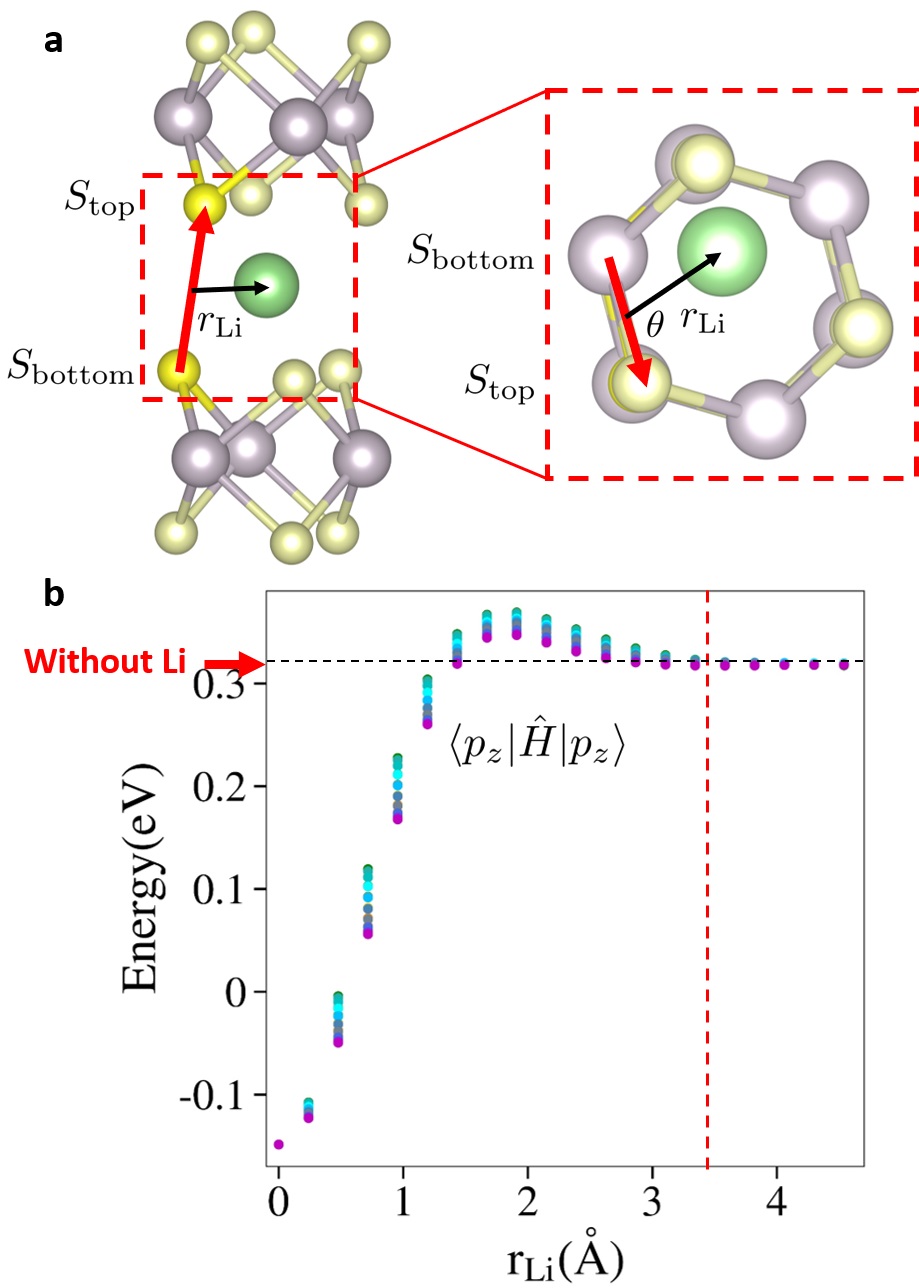}
    \caption{
    \textbf{(a)} Diagram of the atomic sites and relevant variables for the interlayer coupling function under a three-center approximation.
    The red dashed box on the right provides a top-down view. The red arrow indicates the S-S bond, and the angle $\theta$ is measured from the projection of that vector into the $xy$-plane.
    \textbf{(b)} Interlayer $p_z$-$p_z$ S orbital interaction with respect to $\rm r_{Li}$, with different colors corresponding to different values of $\theta$.
    The black dotted horizontal line corresponds to the interaction strength without lithium.
    The red dotted vertical line corresponds to the range of influence of the Li intercalant.}
    \label{fig:3}
\end{figure}

In bilayer MoS$_2$ the interlayer coupling is dominated by interactions between the closest sulfur $p$-orbitals. If there is no lithium present, such couplings have been shown to be well described by the Slater-Koster two-center approximation~\cite{slater1954,fang2015}. However, the introduction of Li atoms at varying distances and orientations will require additional parameters to describe accurately the interaction between S atoms in the two MoS$_2$ layers. We specify the location of a Li atom using polar coordinates $(r_\mathrm{Li},\theta)$, where $r_\mathrm{Li}$ is the distance of the Li atom from the center of the sulfur-sulfur bond of interest, and $\theta$ is angle in the $xy$-plane measured relative to the projection of the S-S bond, as shown in Fig.~\ref{fig:3}. 

From the MLWF basis we can extract the sulfur-sulfur matrix elements as a function of Li position, $(r_\mathrm{Li},\theta)$. In Fig.~\ref{fig:3}b we plot the $p_z$-$p_z$ matrix element as a function of $r_\mathrm{Li}$, with different colors representing different values of $\theta$. It is clear that the presence of Li can have a dramatic effect on the matrix element when Li is nearby, changing both the magnitude and even the sign. When Li is further than $\sim 3$\AA{} from the center of the S-S bond, the effect is negligible. Furthermore, the orientation of the Li atom has a non-negligible but second-order effect. The other combinations of $p$-$p$ matrix elements show similar behavior.

\begin{figure}[htbp]
    \centering
    \includegraphics[width=0.5\linewidth]{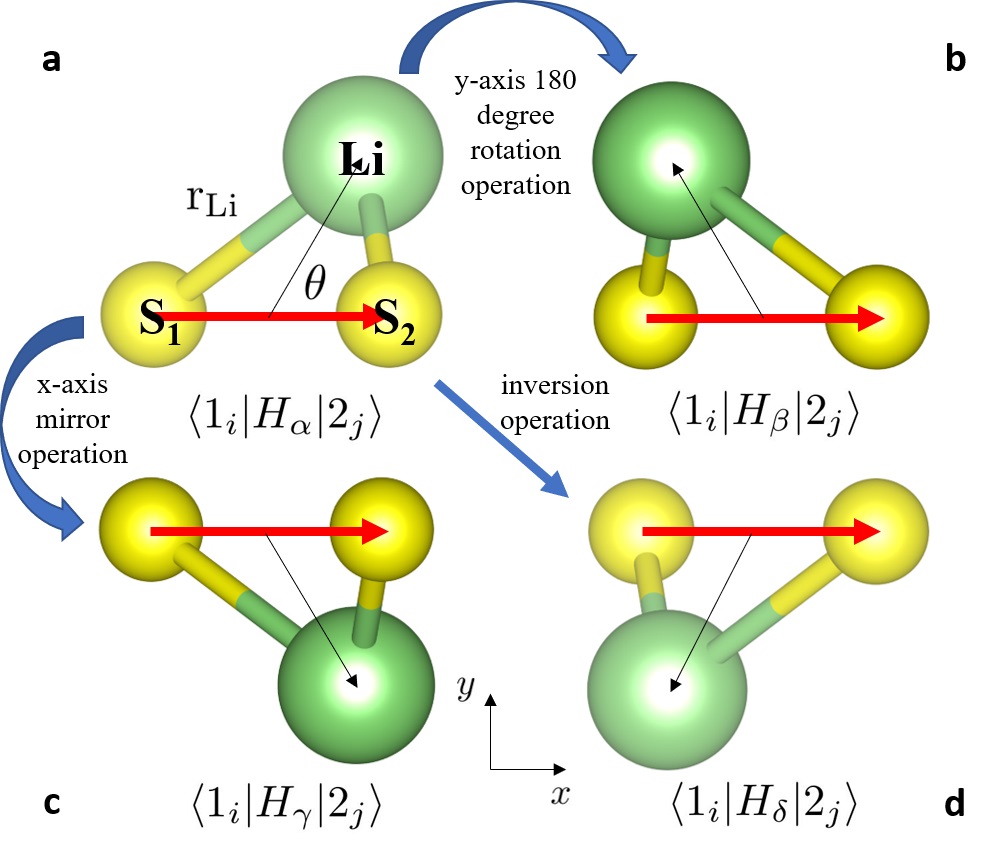}
    \caption{
    Top-down view of four symmetry-related configurations under three symmetry operations.
    \textbf{(a-d)} correspond to $\alpha$, $\beta$, $\gamma$, and $\delta$ configuration characterized by ($r_{\rm Li}$, $\theta$), ($r_{\rm Li}$, $\pi-\theta$), ($r_{\rm Li}$, $-\theta$), and ($r_{\rm Li}$, $\theta-\pi$) respectively.
    The left (right) yellow sphere is the sulfur atom from the bottom (top) layer labelled by 1 (2) and the green sphere denotes the lithium ion.
    $i$ and $j$ denote the atomic $p$ orbitals $p_x$, $p_y$, and $p_z$.
    The $\alpha$ and $\beta$ configurations are related with a rotation by $\pi$ around the $y$-axis, $\alpha$ and $\gamma$ are related by reflection in the $xz$-plane, and $\alpha$ and $\delta$ are related by inversion.
    The $\langle1_i|H_k|2_j\rangle$ ($k$ = \{$\alpha$, $\beta$, $\gamma$, $\delta$\}) denotes the general interlayer sulfur-sulfur $p$-$p$ orbital couplings.}
    \label{fig:tmp}
\end{figure}

We have analyzed how the matrix elements transform under various symmetry operations to constrain the functional form of the interlayer coupling. We considered the following operations as shown in Fig.~\ref{fig:tmp}:
\begin{enumerate}
    \item reflection in the $xz$-plane,
    \item rotation by $\pi$ about the $y$-axis,
    \item inversion (combination of 1 and 2).
\end{enumerate}
For example, consider the relation between $\langle1_x|H_{\alpha}|2_y\rangle$ and $\langle1_x|H_{\gamma}|2_y\rangle$, where lower-case Greek letters refer to a given Li location ($r_\mathrm{Li},\theta$). After applying an $xz$-plane reflection, the $\alpha$ configuration will transform into the $\gamma$ configuration and the S atoms are mapped to themselves: 1$\rightarrow$1, 2$\rightarrow$2, and the $p$-orbitals transform as: $|p_x\rangle\rightarrow|p_x\rangle$, $|p_y\rangle\rightarrow-|p_y\rangle$, and $|p_z\rangle\rightarrow|p_z\rangle$. Thus $\langle 1_x | H_{\alpha} | 2_y\rangle = -\langle 1_x | H_{\gamma} | 2_y\rangle$, or, $t_{xy}(\theta)=-t_{xy}(-\theta)$. Following the same procedure, we can derive the transformations of the couplings under all the three symmetry operations. The results are summarized in Table~\ref{tab:1}.

\begin{table}[htbp]
    \centering
    \begin{tabular}{ P{2cm} P{2cm} P{2cm} P{2cm} }
    \hline\hline
             & $xz$-plane reflection & $y$-axis $\pi$ rotation & inversion \\
    \hline
        $t_{p_x, p_x}$ & $+$ & $+$ & $+$ \\
        $t_{p_x, p_y}$ & $-$ & N/A & N/A \\
        $t_{p_x, p_z}$ & $+$ & N/A & N/A \\
        $t_{p_y, p_x}$ & $-$ & N/A & N/A \\
        $t_{p_y, p_y}$ & $+$ & $+$ & $+$ \\
        $t_{p_y, p_z}$ & $-$ & N/A & N/A \\
        $t_{p_z, p_x}$ & $+$ & N/A & N/A \\
        $t_{p_z, p_y}$ & $-$ & N/A & N/A \\
        $t_{p_z, p_z}$ & $+$ & $+$ & $+$ \\
    \hline
    \end{tabular}
    \caption{Transformation of interlayer sulfur-sulfur $p$-$p$ orbital couplings under reflection, rotation, and inversion. $+$, $-$, and N/A denote not changed, a sign change, and no relation after the given operation.}
    \label{tab:1}
\end{table}

Based on the Slater-Koster two-center approximation for the $p$-$p$ orbital interaction, we can use two functions ($V_{pp,\pi}$ and $V_{pp,\sigma}$) to describe the nine couplings. They satisfy the following relation~\cite{fang2015}:
\begin{equation}
    \begin{aligned}
        V_{pp,\pi}(\textbf{r}) &= \frac{1}{2}\sum_{i}{t_{p_i^{'},p_i}(\textbf{r})}-\frac{1}{2}\sum_{i,j}{t_{p_i^{'},p_j}(\textbf{r})\frac{r_i r_j}{r^2}}\\
        V_{pp,\sigma}(\textbf{r}) &= \sum_{i,j}{t_{p_i^{'},p_j}(\textbf{r})\frac{r_i r_j}{r^2}},
    \end{aligned}
\end{equation}
but in the presence of Li also become functions of $(r_\mathrm{Li}, \theta)$.
Applying the results of the symmetry analysis, the two Slater-Koster functions have the following symmetries for a fixed displacement between the sulfur atoms (fixed $\textbf{r}$):
\begin{equation}
    \begin{array}{rcccl}
        V_{pp, \sigma}(r_{\rm Li},  \theta) &=& V_{pp, \sigma}(r_{\rm Li}, - \theta) &=& V_{pp, \sigma}(r_{\rm Li}, \pi- \theta ) \\
	    V_{pp, \pi}(r_{\rm Li},  \theta) &=& V_{pp, \pi}(r_{\rm Li}, - \theta) &=& V_{pp, \pi}(r_{\rm Li}, \pi+ \theta ).
    \end{array}
\end{equation}
Thus $V_{pp,\sigma}$ and $V_{pp,\pi}$ are completely determined for $\theta$ in the first quadrant. However, even with these constraints the form of the interlayer coupling is still highly complex.
Nonetheless, the MLWF procedure captures the microscopic details of how Li affects the interlayer coupling.

\section{Effect of Lithium on the moir\'e potential}\label{potential}

The tight-binding approach demonstrates the dramatic influence that Li intercalants can have on interlayer couplings, but a full model accurately incorporating all the additional degrees of freedom introduced by a Li atom would be extremely complicated. To understand the effects that Li can have on the interlayer interactions in a twisted cell, we turn to a simpler continuum model for twisted bilayer TMDCs \cite{wu2018,wu2019}. In contrast to the empirical form of interlayer interactions that have been used to study bilayer graphene~\cite{rafi2011}, here we use \textit{ab initio} calculations to accurately obtain the interlayer moir\'e potential by careful study of the DFT band structure \cite{Jung2014}. In our case, we will focus on the twisted 2H-bilayer MoS$_2$  

Because Li intercalation raises the Fermi level into the conduction bands of MoS$_2$, we will construct a continuum model for the electronic states near the conduction band edge which can be described using the effective mass approximation.
The two lowest parabolic conduction bands at the $K$ point are a pair of identical bands from each TMDC layer, with an energy splitting caused by interlayer hybridization. Note that for the 2H orientiaton, the $K$ point of the bottom layer corresponds to the $K'$ point of the top layer.
However, since we can safely ignore spin at the conduction $K$ points, $K$ and $K'$ give identical band edges.
For a twisted bilayer, a simple but robust model is comprised of two monolayer bands with effective mass $m^*$ and an interlayer coupling $V(\vec{r})$, taken to be a smooth function of the position in the moir\'e supercell:

\begin{equation}
    H_{\vec{k}} = 
    \begin{pmatrix}
    - \frac{\hbar \vec{k}^2}{2 m^*} & V(\vec{r}) \\
    V^\dagger(\vec{r}) & - \frac{\hbar \vec{k}^2}{2 m^*}
    \end{pmatrix}
\end{equation}

A more careful consideration of the problem also takes into account a stacking-dependent onsite energy for both monolayer bands~\cite{wu2019}.
More specifically, this captures not just the band splittings but also the bands' average energy.
In contrast to the explicit interlayer coupling $V(\vec{r})$, this onsite term represents changes to the in-plane electronic structure due to the presence of the neighboring layer, and so also depends on the stacking.
Here we ignore this contribution, since our focus is on how lithium intercalation modifies interlayer hybridization to facilitate the appearance of flat-bands, but such a term can play an important role in the complete model.

\begin{figure*}[htbp]
    \centering
    \includegraphics[width=\linewidth]{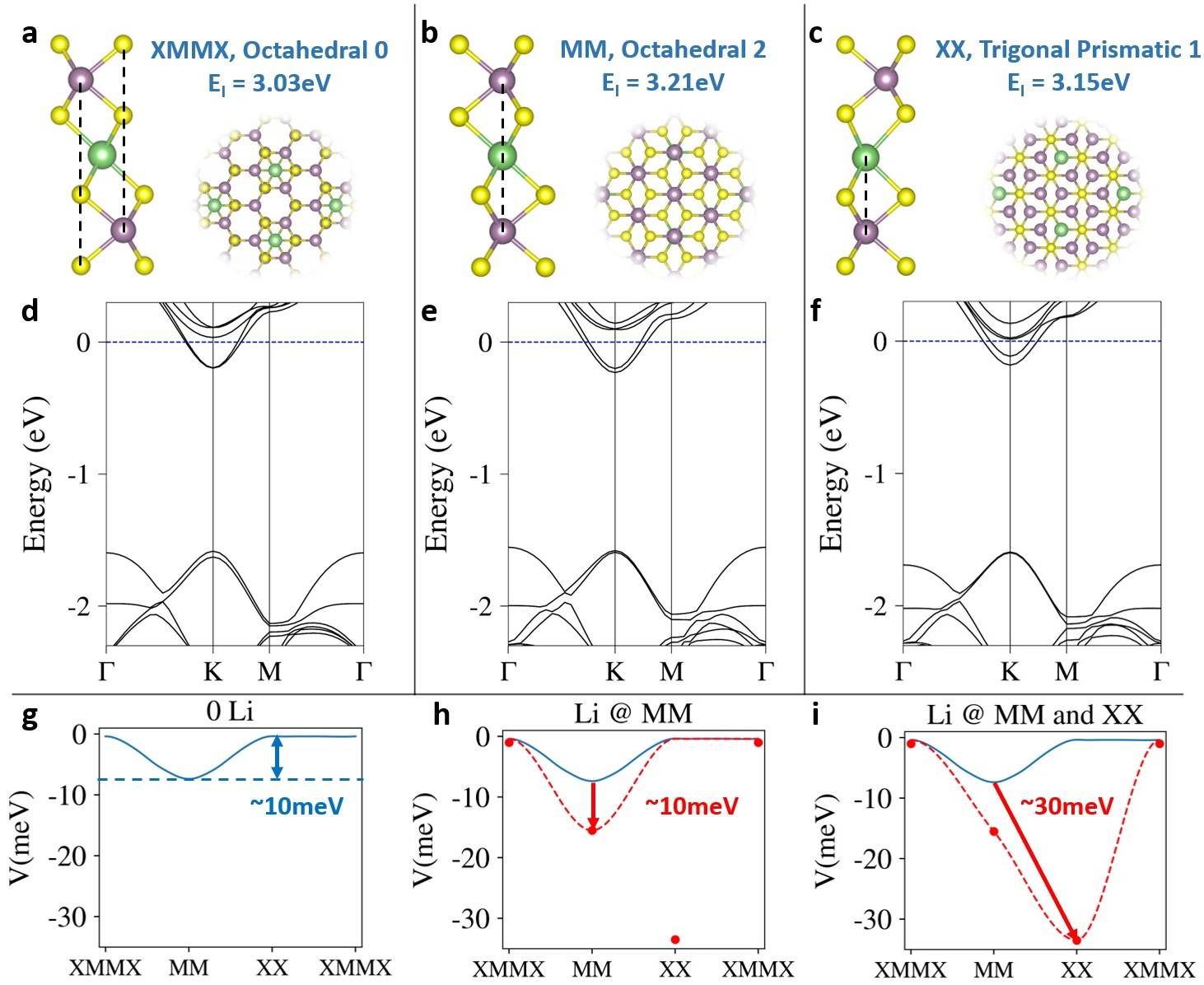}
    \caption{
    Effects of Li on the moir\'e potential in 2H-bilayer MoS$_2$.
    \textbf{(a-c)} Configurations with the most favorable intercalation energy ($E_{\rm I}$): Li in the XMMX, MM, and XX regions with an intercalation density of 1 lithium per 2$\times$2 supercell.
    \textbf{(d-f)} Band structures corresponding to the configurations shown in \textbf{(a-c)}.
    \textbf{(g-i)} Real space distribution of the interlayer moir\'e potential along the diagonal direction of the moir\'e supercell. From left to right: without lithium, with lithium only in the MM region, and with lithium in both the MM and XX region.}
    \label{fig:4}
\end{figure*}

The interlayer interaction $V(\vec{r})$ acts like a potential energy for the electronic states in the moir\'e supercell, thus we refer to it as the interlayer moir\'e potential.
From the variation in this potential throughout the moir\'e supercell we can obtain the band structure for the twisted bilayer.
The potential $V(\vec{r})$ across the entire moir\'e supercell can be estimated from its value at the three high-symmetry stacking arrangements.
The magnitude of $V(\vec{r})$ is simply one half of the energy splitting of the lowest two states at the $K$ conduction band edge (Fig.~\ref{fig:4}), while its phase is determined by details of the atomic geometry.
Because the conduction band edge has primarily $d$ orbital character, $V(\vec{r})$ is largest in the non-intercalated case when the metal atoms are aligned (MM configuration). 
Taking the MM stacking to correspond to $\vec{r} = 0$ and expanding the interlayer potential in the lowest harmonics of the reciprocal lattice vectors for the moir\'e supercell, $\vec{G^{sc}}_i$, gives~\cite{rafi2011,wu2019}

\begin{equation}
V(\vec{r}) = \frac{V_0}{3} \left( 1 + e^{-i \vec{G^{sc}_1} \cdot \vec{r}} + e^{-i \vec{G^{sc}_2} \cdot \vec{r}} \right).
\end{equation}

Compared to  the results of Wu et al~\cite{wu2019},  who studied TMDC homobilayers in an aligned phase, our system is in the non-aligned 2H phase.
Although this provides a different crystal symmetry the expansion is still a good match in the unintercalated case, as the splitting of the conduction band-edge at $K$ mostly depends on the distance between the metal atoms and not the detailed symmetry of the crystal structure.
This enters the $\vec{k}$-dependent hamiltonian $H_{\vec{k}}$ as three interlayer momentum scattering terms of equal strength~\cite{rafi2011}.
The prefactor above is chosen as $V_0/3$ such that $V(0) = V_0$.
Note that at the other two high-symmetry stackings, $\vec{r} = n(\vec{a^{sc}}_1 + \vec{a^{sc}}_2)/3$ for $n=1,2$, the function $V$ is identically zero.
Lying along the diagonal direction of the moir\'e supercell, these are the same three local stacking regions which provide stable intercalation sites for lithium.
The most stable intercalation site for each of these local configurations is shown in Fig.~\ref{fig:4}a-c, with the band structure for a uniform cell with that stacking configuration immediately below in Fig.~\ref{fig:4}d-f.

Combining the results for the band splitting in each different stacking configuration, we can assemble the moir\'e potentials shown in Fig.~\ref{fig:4}g-i.
The blue line in all three panels shows the moir\'e potential along the diagonal direction of the moir\'e supercell in 2H-bilayer MoS$_2$ without lithium intercalation.
It has a potential well around the MM stacking configuration which can localize the electronic states and presumably lead to flat bands and correlated physics.
However, the potential well is only $\sim$10 meV deep.

Introducing Li between the layers, which will first condense around the MM stacking regions because they offer the most energetically favorable intercalation sites, leads to a deeper moir\'e potential near the MM stacking by a factor of 2. This makes flat bands and correlated interactions more likely. Further increase of the intercalation density results in lithium accumulation in XX stacking regions, deepening the moir\'e potential but also changing its shape. 
Moving the origin ($\vec{r} = 0$) of the moir\'e potential amounts to a gauge choice in the interlayer coupling that does not affect the moir\'e band structure in the continuum model~\cite{rafi2011}.
However, the lowest harmonic approximation of $V(\vec{r})$ now fails as the values at MM and XMMX are not identical.
Since we only aim to estimate the effect of lithium intercalation on the moir\'e band structure we ignore this complication and treat the potential well as for the MM case.
To calculate the moir\'e band structure more accurately, it is necessary to include higher harmonics in the interlayer potential.
Obtaining these harmonics from DFT calculations requires band structures for additional Li intercalated geometries, which can be challenging to optimize for low-symmetry stacking configurations.

\section{Emerging flat-bands}\label{flat-bands}

\begin{figure*}[htbp]
    \centering
    \includegraphics[width=\linewidth]{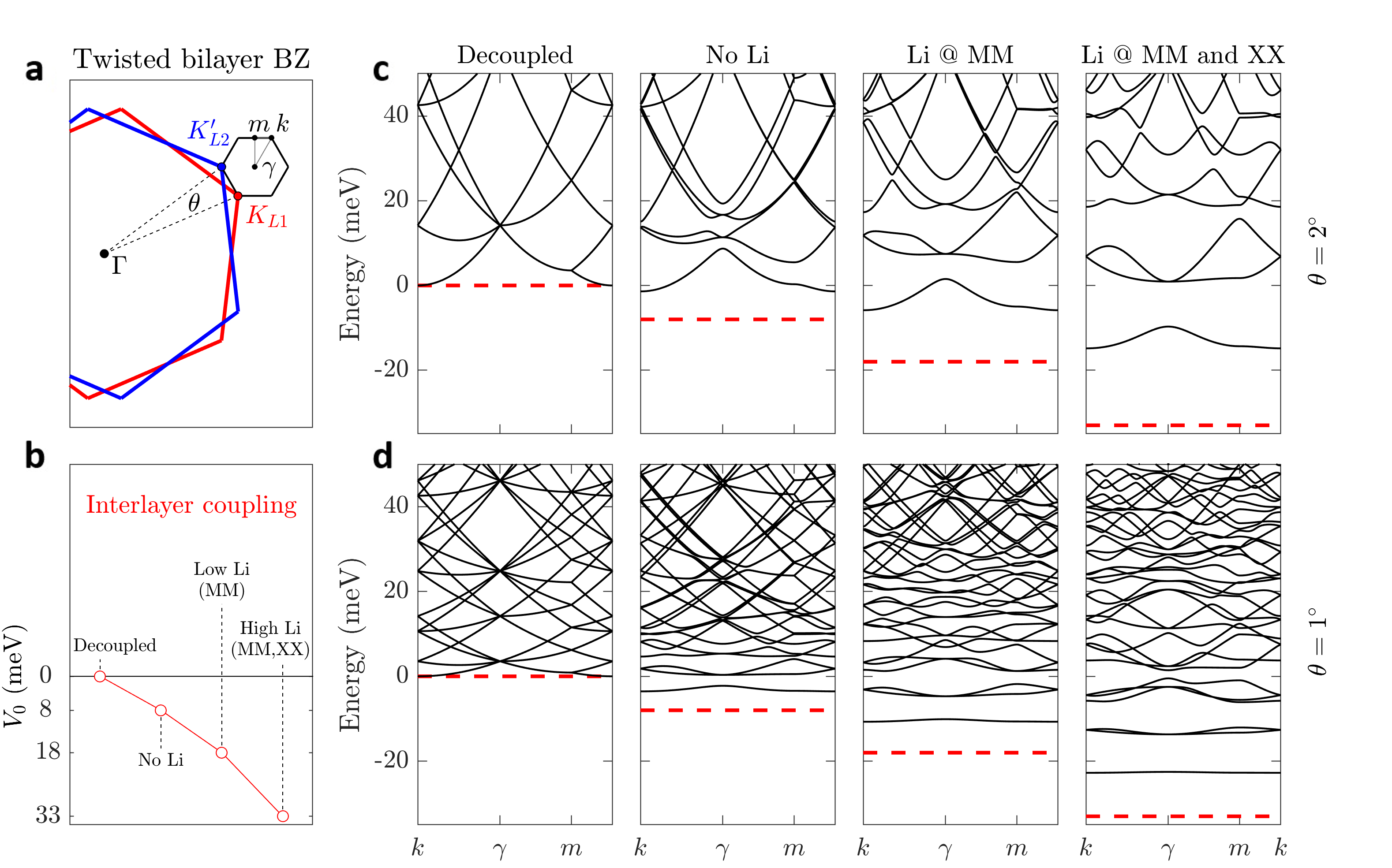}
    \caption{
    Emerging flat bands in twisted 2H-bilayer MoS$_2$ with lithium intercalation. \textbf{(a)} Brillouin zone (BZ) of the bottom layer (red) and top layer (blue) of a twisted 2H-bilayer MoS$_2$.
    The moir\'e superlattice BZ is given in black, with its high-symmetry points in lowercase labels.
    \textbf{(b)} Dependence of the interlayer coupling strength on lithium intercalation in 2H-bilayer MoS$_2$.
    A case with no interlayer coupling (decoupled) is included along with the three configurations of Li intercalants studied.
    The axis is chosen to help comparison with band structures.
    \textbf{(c,d)} Band structures of twisted 2H-bilayer MoS$_2$ for $\theta$ equal to $2^\circ$ and $1^\circ$ using the continuum model.
    The interlayer coupling strength is varied by the details of lithium intercalation.
    The bands here are expanded from the conduction $K$ valley, which is most relevant because of electron doping by the Li intercalants.
    The band edge of the monolayer conduction band is taken as $0$ energy and the dashed red-line shows the maximum depth of the interlayer-coupling potential in each case.}
    \label{fig:5}
\end{figure*}

The strong effect of lithium intercalation on the interlayer moir\'e potential is useful in enhancing the flat bands of a twisted TMDC bilayer.
In Fig.~\ref{fig:5} the band structure obtained from the continuum model for the $K$ point conduction bands of twisted 2H-bilayer MoS$_2$ are shown for two different twist angles ($2^\circ$, $1^\circ$) and the three cases of Li intercalation (zero, Li only at MM, and Li at both MM and XX).
To match the DFT band structure calculations the value of $V_0$ for three lithium intercalation cases is taken to be 8, 18, and 33 meV, respectively.
The effective mass for the conduction band edge at $K$ is obtained by fitting to the monolayer band structure, yielding $m^*/\hbar = 75 $ meV$^{-1}$ \AA{}$^{-2}$.

For a $2^\circ$ twist angle, the unintercalated bilayer does not have a splitting between the lowest two superlattice bands, and is thus very far from flat band behavior.
With Li intercalation the strength of the moir\'e potential can be increased by a factor of 4, allowing flat bands to emerge even at this large twist angle.
The $1^\circ$ system has flat bands visible even at zero intercalation, but the intercalation greatly increases the gap between the superlattice bands and flattens the bands further. As Li concentration increases, not only does the lowest conduction band become flatter, but the bottom of the moir\'e potential well shifts from the regions of MM stacking to regions with XX stacking.
As the moir\'e flat-bands host large ``spots'' of electronic density confined by this potential well, a shifting of the electron density from MM stacking to XX stacking would be visible, for instance, in scanning tunneling microscopy.

\section{Conclusion}\label{conclusion}
We have performed first-principles DFT calculations and \textit{ab initio} tight-binding and continuum modeling to explore the structural and electronic properties of lithium intercalation in twisted bilayer MoS$_2$ systems. We found that lithium intercalants have the lowest energy in MM and XX stacking regions in the moir\'e supercell. The presence of Li dramatically enhances the interlayer interaction, increasing the depth of the effective moir\'e potential well from $\sim$8 meV to $\sim$20 meV for Li in MM regions, and further to $\sim$33 meV for Li in XX stacking regions. 
Using a continuum model with an interlayer interaction based on these DFT results, we show that lithium intercalation can better flatten and isolate the conduction bands near the Fermi level. Furthermore, such moir\'e flat bands can be realized at larger twist angles with the aid of intercalation. Our results demonstrate that intercalation can be a powerful tool for controlling moir\'e flat bands in twistronic devices.

\begin{acknowledgments}

We acknowledge Yiqi Xie for helpful discussions. 
The computations in this paper were run on the FASRC Odyssey cluster supported by the FAS Division of Science Research Computing Group at Harvard University.
S.C. was supported by ARO MURI Award No. W911NF-14-0247 and by the STC Center for Integrated Quantum Materials, NSF Grant No. DMR-1231319.

\end{acknowledgments}

\newpage
\bibliography{ref}
\end{document}